\documentclass[english, reprint, aps, prb, twocolumn, superscriptaddress, showpacs, floatfix]{revtex4-1}
\usepackage{babel}
\usepackage{amssymb}
\usepackage{mathrsfs} 
\usepackage{amsmath}
\usepackage{float}
\usepackage{bbm}
\usepackage{graphicx}
\usepackage{epsfig}
\usepackage{epstopdf}
\usepackage{verbatim}
\usepackage{multirow}
\usepackage[usenames]{color}

\usepackage{hyperref}
\begin{document}

\title{Benchmarking of dynamically corrected gates for the exchange-only spin qubit in $1/f$ noise environment}
\author{Chengxian Zhang}
\affiliation{Department of Physics and Materials Science, City University of Hong Kong, Tat Chee Avenue, Kowloon, Hong Kong SAR, China}
\author{Xu-Chen Yang}
\affiliation{Department of Physics and Materials Science, City University of Hong Kong, Tat Chee Avenue, Kowloon, Hong Kong SAR, China}
\author{Xin Wang}
\email{Correspondence author, x.wang@cityu.edu.hk}
\affiliation{Department of Physics and Materials Science, City University of Hong Kong, Tat Chee Avenue, Kowloon, Hong Kong SAR, China}
\date{\today}

\begin{abstract}
We study theoretically the responses of the dynamically corrected gates to time-dependent noises in the exchange-only spin qubit system. We consider $1/f$ noises having spectra proportional to $1/\omega^\alpha$, where the exponent $\alpha$ indicates the strength of correlation within the noise. The quantum gate errors due to noises are extracted from a numerical simulation of Randomized Benchmarking, and are compared between the application of  uncorrected operations and that of dynamically corrected gates robust against the hyperfine noise. We have found that for $\alpha\gtrsim1.5$, the dynamically corrected gates offer considerable reduction in the gate error and such reduction is approximately two orders of magnitude for the experimentally relevant noise exponent. On the other hand, no improvement of the gate fidelity is provided for $\alpha\lesssim1.5$. This critical value $\alpha_c\approx1.5$ is comparatively larger than that for the cases for the singlet-triplet qubits. The filter transfer functions corresponding to the dynamically corrected gates are also computed and compared to those derived from uncorrected pulses. Our results suggest that the dynamically corrected gates are useful measures to suppress the hyperfine noise when operating the exchange-only qubits.
\end{abstract}


\maketitle

\section{introduction}

Spin qubits confined in semiconductor quantum dots are promising candidates for quantum computing\cite{Loss.98,Taylor.05} due to their demonstrated long coherence time, high control fidelities\cite{Petta.05, Bluhm.10b, Barthel.10, Maune.12, Pla.12, Pla.13, Muhonen.14, Kim.14,Kawakami.16} as well as expected scalability. A natural approach, proposed by Loss and DiVincenzo,\cite{Loss.98} is to encode one qubit using the spin up and down states of a single electron. However, difficulties in performing ESR-type single electron spin rotations\cite{Koppens.06} in this kind of the qubit have led researchers to propose alternative ways to encode qubits in the collective states of two or more electrons.\cite{Levy.02} The singlet-triplet qubit is the simplest qubit that can be controlled all-electrically via the exchange interaction,\cite{Petta.05,Maune.12} but its full control still requires a magnetic field gradient.\cite{Foletti.09,Bluhm.10c,Brunner.11,Petersen.13,Wu.14} In a seminal paper,\cite{DiVincenzo.00} DiVincenzo proposed a qubit employing certain three-spin states, which can be controlled solely by the exchange interaction, and is thus termed as the ``exchange-only'' qubit. The exchange-only qubit, together with its variant---the resonant exchange qubit, has been experimentally demonstrated at the single-qubit level.\cite{Laird.10,Gaudreau.12,Medford.13,Eng.15} Nevertheless, hyperfine-mediated nuclear spin fluctuations\cite{Medford.13b,Hung.14} as well as charge noises\cite{Fei.15,Russ.15,Shim.16} contribute to decoherence, preventing the implementation of more complicated operations required to operate two or more qubits, despite extensive theoretical studies on the two qubit gates.\cite{DiVincenzo.00,Fong.11,Taylor.13,Doherty.13,Setiawan.14,Zeuch.14,Arijeet.15,Wardrop.16, Zeuch.16}
A comprehensive understanding of the interaction between noises and controls is therefore of crucial importance for the field to progress.

Dynamically corrected gates\cite{Khodjasteh.09,Khodjasteh.10, Khodjasteh.12, Wang.12, Green.12, Kosut.13}  (DCGs) are useful measures to combat decoherence. Inspired by the dynamical decoupling technique developed in the field of NMR quantum control,\cite{Hahn.50, Carr.54, Meiboom.58, Uhrig.07, West.12} DCGs have been successfully developed to reduce both hyperfine and charge noise in the singlet-triplet qubit\cite{Wang.12,Kestner.13,Wang.14a,Wang.14b,Wang.15} as well as the exchange-only qubit.\cite{Hickman.13,Setiawan.14}, i.e. the noises are assumed to vary with a much longer time scale than typical gate operations. With this assumption, the DCGs are tailored under the static noise model, usually by canceling the effect of noise on the evolution operator up to certain orders using piecewise constant pulses. In realistic situations, the DCGs should work well for the low frequency components of the noises but not high frequency parts. Theoretical validation of this approximation has been performed through the Randomized Benchmarking\cite{Knill.08,Magesan.12} for the singlet-triplet qubits\cite{Wang.14a,Wang.14b, Yang.16} under the $1/f$ noise,\cite{kogan} the power spectral density of which is proportional to $1/\omega^\alpha$. It has been shown there that the DCGs offer great error reduction for $\alpha\gtrsim\alpha_c$ but no error cancellation otherwise, where for the DCGs developed for the singlet-triplet qubit system\cite{Kestner.13,Wang.14a,Wang.14b,Yang.16} the critical $\alpha_c\approx1$. The validity of DCGs can then be assessed by measuring the noise spectra before they are actually being implemented, the feature of which is very useful for their experimental realization because the noise spectra are typically easier to be mapped out\cite{Rudner.11,Medford.12,Eng.15} than actually carrying out the relatively complicated composite sequences.\cite{Rong.14}

On the other hand, however, benchmarking of DCGs for the exchange-only qubits under realistic noises has been lacking in the literature. This is an important open question because there are considerable differences between the exchange-only qubits and singlet-triplet qubits. Firstly, the pulse sequences now involve two exchange interactions thus are more complicated.\cite{Hickman.13} Secondly, the hyperfine noise not only causes dephasing as in other types of spin qubits,\cite{Petta.05, Maune.12} but also leads to leakage outside of the computational subspace.\cite{Ladd.12} Third, the pulse sequences are longer than those of the singlet-triplet qubit.\cite{Hickman.13} One may therefore speculate that the sequences proposed in Ref.~\onlinecite{Hickman.13} would require noises with a larger $\alpha_c$ compared to ones for singlet-triplet qubits, namely the noise must be more correlated for the DCGs to work in the exchange-only qubit.
It is unknown, without any quantitative results, that whether $\alpha_c$ would exceed the experimentally measured value, for example $\alpha\approx2.6$ (cf. Ref.~\onlinecite{Rudner.11,Medford.12}), rendering the DCGs useless. 
It is therefore an important problem to study the effect of DCGs undergoing realistic noise for the exchange-only qubit system, in particular the determination of $\alpha_c$, which implies the range of the noise spectra with which the DCGs work.

In this paper, we perform a numerical study on how DCG pulses perform under $1/f$ noises for the exchange-only qubit. We focus on the sequences developed in Ref.~\onlinecite{Hickman.13} which correct the hyperfine noise. We numerically simulate the Randomized Benchmarking,\cite{Knill.08,Magesan.12} comparing sequences composed of uncorrected and corrected single-qubit Clifford gates respectively. We find that the critical noise exponent for the DCGs developed in Ref.~\onlinecite{Hickman.13} is $\alpha\approx1.5$. Although this value is larger than that found for the \textsc{supcode} sequences of singlet-triplet qubits, it is still lower than the experimentally measured ones available in the literature,\cite{Rudner.11,Medford.12} reaffirming the noise-compensating power of the DCG sequences. We also discuss the filter transfer functions\cite{Green.12, Green.13,Paz-Silva.14,Ball.16}, which offer complementarily useful information to the benchmarking.\cite{Kabytayev.14,Ball.16}

The remainder of this paper is organized as follows. In Sec.~\ref{sec:model} we give a summary of the model of an exchange-only qubit, the noise involved, and the DCGs that we are going to employ in this work. Sec.~\ref{sec:results} presents our results, including the Randomized Benchmarking, fidelity decay constant, the improvement ratio of the DCGs and the filter transfer functions. We conclude in Sec.~\ref{sec:conclusion}.

\section{Model}\label{sec:model}

\begin{figure}
  (a)\includegraphics[width=0.45\columnwidth]{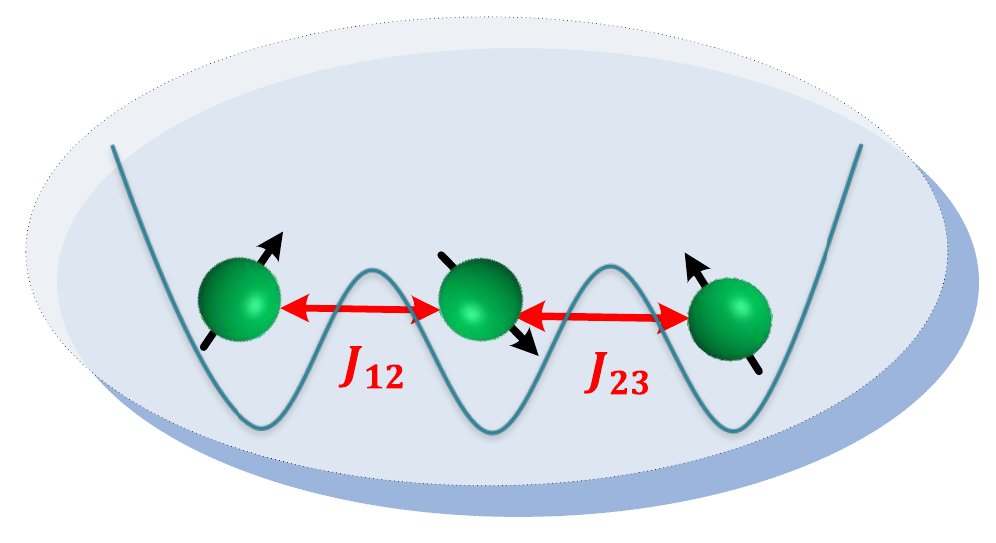}
  (b)\includegraphics[width=0.9\columnwidth]{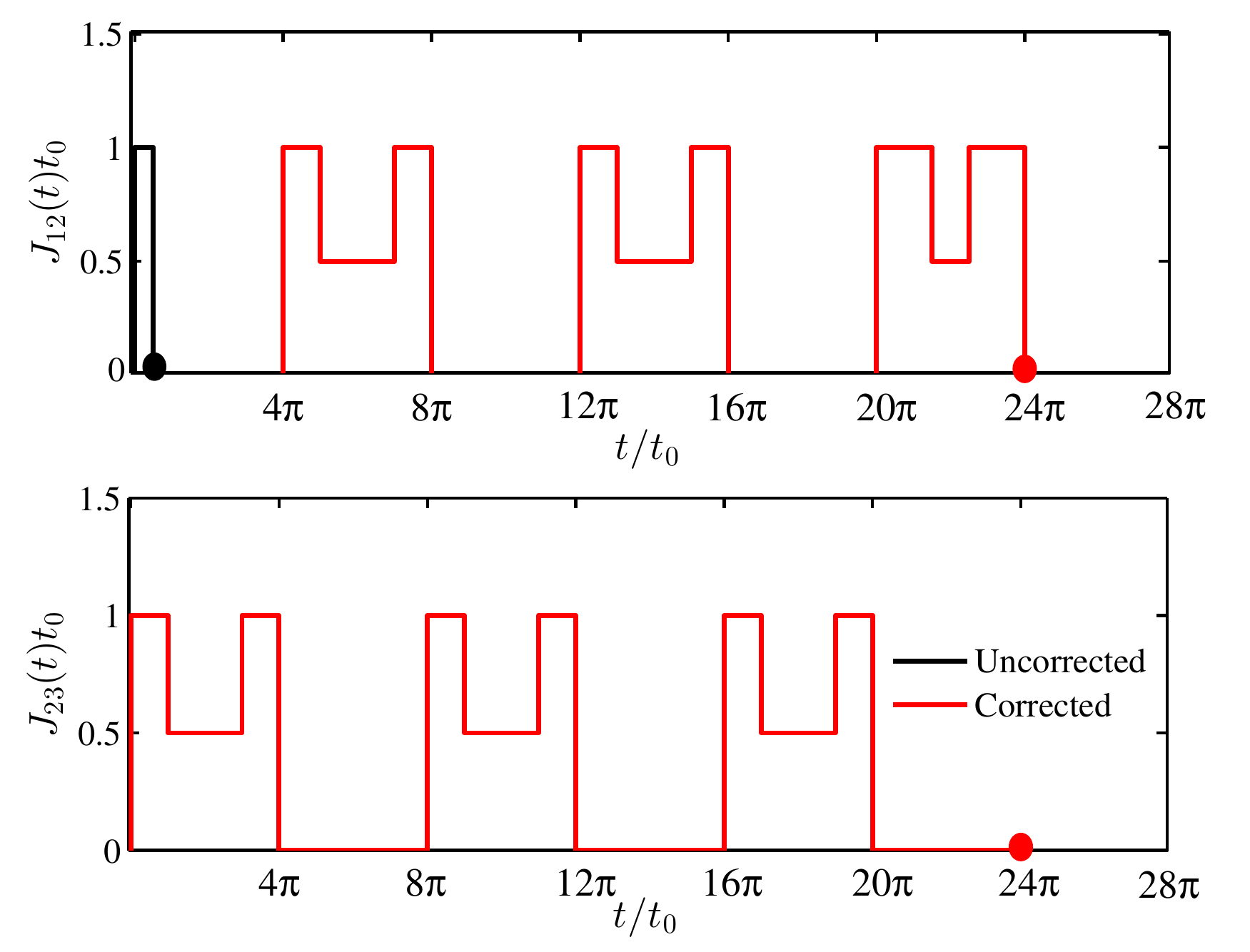}
\caption{(a) A schematic of an exchange-only qubit with exchange interactions between adjacent spins indicated. (b) Example of the dynamically corrected gates achieving a rotation of $R_{12}(\pi/2)$, canceling the leading order contribution from the hyperfine noise. $t_0$ is an arbitrary time unit.}
\label{fig:pulseexample}
\end{figure}

The exchange-only qubit is encoded in the $S=1/2$ and $S^z=1/2$ subspace of the three-electron system as $|0\rangle=({|\!\uparrow\downarrow\uparrow\rangle}-{|\!\downarrow\uparrow\uparrow\rangle})/\sqrt{2}$ and $|1\rangle=({|\!\uparrow\downarrow\uparrow\rangle}+{|\!\downarrow\uparrow\uparrow\rangle})/\sqrt{6}-\sqrt{6}{|\!\uparrow\uparrow\downarrow\rangle}/3$ (cf. Ref.~\onlinecite{DiVincenzo.00}). Inhomogeneous fluctuations in the magnetic field cause the qubit states to leak to an $S=3/2$, $S^z=1/2$ state $|Q\rangle=({|\!\uparrow\downarrow\uparrow\rangle}+{|\!\downarrow\uparrow\uparrow\rangle}+{|\!\uparrow\uparrow\downarrow\rangle})/\sqrt{3}$ via the hyperfine coupling $H_{\rm hf} = \sum_j B_jS^z_j$, where $S^z_j$ is the spin operator in the $z$-direction for the $j$th electron, and $B_j$ the hyperfine field.\cite{Ladd.12} The Hamiltonian of an exchange-only qubit can then be written under the basis $\{|0\rangle, |1\rangle, |Q\rangle\}$. It contains two parts, $H=H_{\rm c}+H_{\rm hf}$, each of which can be expressed in terms of the Gell-Mann matrices,\cite{Georgi.99} $\lambda_j$, as\cite{Ladd.12}
\begin{align}
H_{\rm c}=J_{12}(t)E_{12}+J_{23}(t)E_{23},\label{eq:ham}
\end{align}
with
\begin{align}
E_{12}=-\frac{\lambda_3}{2}-\frac{\lambda_8}{2\sqrt{3}}, \quad E_{23}=-\frac{\sqrt{3}}{4}\lambda_1+\frac{\lambda_3}{4}-\frac{\lambda_8}{2\sqrt{3}},\label{eq:E12E23}
\end{align}
and
\begin{align}
H_{\rm hf}=\left(\frac{\lambda_1}{2 \sqrt{3}}+\frac{\lambda_4}{\sqrt{6}}\right)\Delta_A+\left(\frac{\lambda_3}{3}+\frac{\sqrt{2}}{3}\lambda_6\right)\Delta_B.\label{eq:Hhfdef}
\end{align}
Here, $\Delta_A=B_1-B_2$ and $\Delta_B=B_3-(B_1+B_2)/2$ signify the inhomogeneity of the magnetic field stemming from the hyperfine fluctuations, which have been denoted as $\Delta_{12}$ and $\Delta_{\overline{12}}$ in previous literature.\cite{Ladd.12,Hickman.13}
$J_{ij}(t)$ is the exchange interaction between electrons in dots $i$ and $j$ which are controlled electrostatically [a schematic picture is shown in Fig.~\ref{fig:pulseexample}(a)]. In the absence of leakage, $E_{12}$ and $E_{23}$ implement rotations on the Bloch sphere around two non-perpendicular axes, $\hat{z}$ and $\frac{\sqrt{3}}{2}\hat{x}-\frac{1}{2}\hat{z}$, respectively. These ideal rotations are denoted as $R_{12}(\phi)=\exp(-iE_{12}\phi)$ and $R_{23}(\phi)=\exp(-iE_{23}\phi)$, which can then be utilized to implement arbitrary single-qubit rotations.

Operations of an exchange-only qubit are subject to two major channels of noises. The nuclear noise, $H_{\rm hf}$, causes dephasing within the logical subspace and leakage to state $|Q\rangle$. The charge noise, on the other hand, results in errors of the control. Dynamically corrected gates are developed in order to combat these noises and improve the fidelity of the gate control, with the fundamental idea being the ``self-compensation'' of error. In particular, for the exchange-only qubits, it has been shown that cancellation of the leading order error arising from the hyperfine noise can be achieved by a composite pulse sequence containing 18 pieces. Further, if simultaneous resilience of charge noise and hyperfine noise is desired, a longer composite pulse consists of 13 properly combined hyperfine-noise-corrected sequence can cancel both noises to their leading orders.\cite{Hickman.13} These composite pulses are theoretically tailored based on a complete cycle of permutation between different spins, the symmetry property of which has greatly simplified the problem.

An important assumption behind the family of dynamically corrected gates is the. While this quasi-static-noise approximation is justified by the fact that the time scale with which the noise varies is much longer than the typical gate operation time, it remains necessary to benchmark these gate sequences in a more realistic time-dependent noise environment, for example the $1/f$ noise.\cite{kogan} This practice has been done for the \textsc{supcode} sequence developed for the singlet-triplet qubit,\cite{Wang.14a,Wang.14b, Yang.16} and the conclusion is that the DCGs will generally offer improvement for the noise exponent $\alpha\gtrsim1$ but not otherwise. However, the benchmarking remains to be done for the exchange-only qubit because there are important differences. As mentioned above, the pulse sequences for the exchange-only qubit involve two-axis exchange control and are longer than \textsc{supcode}. Moreover, there is one additional channel of error -- the leakage -- which needs to be treated. It is therefore conceivable that the critical noise exponent, $\alpha_c$ may be higher for the DCG of the exchange-only qubit than the \textsc{supcode}. A quantitative evaluation of this statement is one of the main purposes of this paper.

In order to clarify this issue, we numerically performed the Randomized Benchmarking of the 24 single-qubit Clifford gates. In this work, we focus on the hyperfine-noise-corrected sequence only.\cite{Hickman.13} The sequences which cancel both hyperfine and charge noise are much longer\cite{Hickman.13} (which needs rotation around the Bloch sphere by a total angle of about $\sim300\pi$) and is unlikely to provide substantial error reduction compared to the hyperfine-noise-corrected one for similar noises. Further optimization is required for this type of sequences. To facilitate the discussion, we explicitly give the pulse sequence corresponding to a net rotation of $R_{12}(\phi)$ $(-\pi\le\phi\le\pi)$ which cancels hyperfine noise as follows. The sequence, denoted by $\widetilde{U}_{12}(\phi)$, can be expressed as
\begin{multline}
\widetilde{U}_{12}(\phi)\equiv U_{12}'(J,\pi+\phi)U_{23}'(J,\pi)\left[U_{12}'(J,\pi)U_{23}'(J,\pi)\right]^2
\\
=R_{12}(\phi) + \mathcal{O}\left[\left(\Delta_A+\Delta_B\right)^2\right],\label{eq:hfcorrdU12}
\end{multline}
where $U'_{12/23}(J,\phi)$ is defined as
\begin{equation}
U'_{12/23}(J,\phi)=U_{12/23}(J,\phi)U_{12/23}\left(\frac{J}{2},2\pi-\phi\right)U_{12/23}(J,\phi),\label{eq:3piecePhi}
\end{equation}
and $U_{12/23}(J,\phi)$ is the evolution according to the Hamiltonian Eq.~\eqref{eq:ham} under the hyperfine noise
\begin{equation}
U_{12/23}(J,\phi)=\exp\left\{-i\left(J E_{12/23}+H_{\rm hf}\right)\frac{\phi}{J}\right\}.\label{eq:U12def}
\end{equation}
An example of the composite pulse sequence for a rotation $R_{12}(\pi/2)$ is shown in Fig.~\ref{fig:pulseexample}(b). 
For the corresponding sequence of $\widetilde{U}_{23}(\phi)$, one simply interchanges the indices 12 and 23 in Eq.~\eqref{eq:hfcorrdU12}.

We now explain how we construct the 24 single-qubit Clifford gates using the available rotations around the $z$-axis and another $120^\circ$ apart from it. Note that arbitrary single-qubit rotation of angle $\phi$ around the unit vector $\hat{r}$, $R(\hat{r},\phi)$, can be decomposed into rotations around the $x$ and $z$ axis as $R(\hat{z},\phi_a)R(\hat{x},\phi_b)R(\hat{z},\phi_c)$ [cf. Ref.~\onlinecite{Wang.14a}] and $R(\hat{z},\phi)=R_{12}(-\phi)$. In order to express $R(\hat{x},\phi)$ in terms of $R_{12}$ and $R_{23}$, we first decompose $R_{23}(\psi)$ in the same way:
\begin{equation}
R_{23}(\psi)=R(\hat{z},-\phi_a')R(\hat{x},\phi)R(\hat{z},-\phi_c'),
\label{eq:R23decomp}
\end{equation}
finding the exact values of $\phi_a'$ and  $\phi_c'$. Then, $R(\hat{x},\phi)$ is simply
\begin{equation}
R(\hat{x},\phi)=R_{12}(-\phi_a')R_{23}(\psi)R_{12}(-\phi_c'),
\label{eq:R23res}
\end{equation}
where
\begin{align}
\phi_a'&=-\mathrm{Arg}\left(\sqrt{1-\frac{4}{3}\sin^2\frac{\phi}{2}}-\frac{i}{\sqrt{3}}\sin\frac{\phi}{2}\right),\\
\psi&=-2\arcsin\left(\frac{2}{\sqrt{3}}\sin\frac{\phi}{2}\right)\label{eq:psi},
\end{align}
and $\phi_c'=\phi_a'$.

A complete list of the single-qubit Cliiford gates expressed in terms of $R_{12}$ and $R_{23}$ rotations are presented in the Appendix.

\section{Results}\label{sec:results}

\begin{figure}
  \includegraphics[width=0.7\columnwidth]{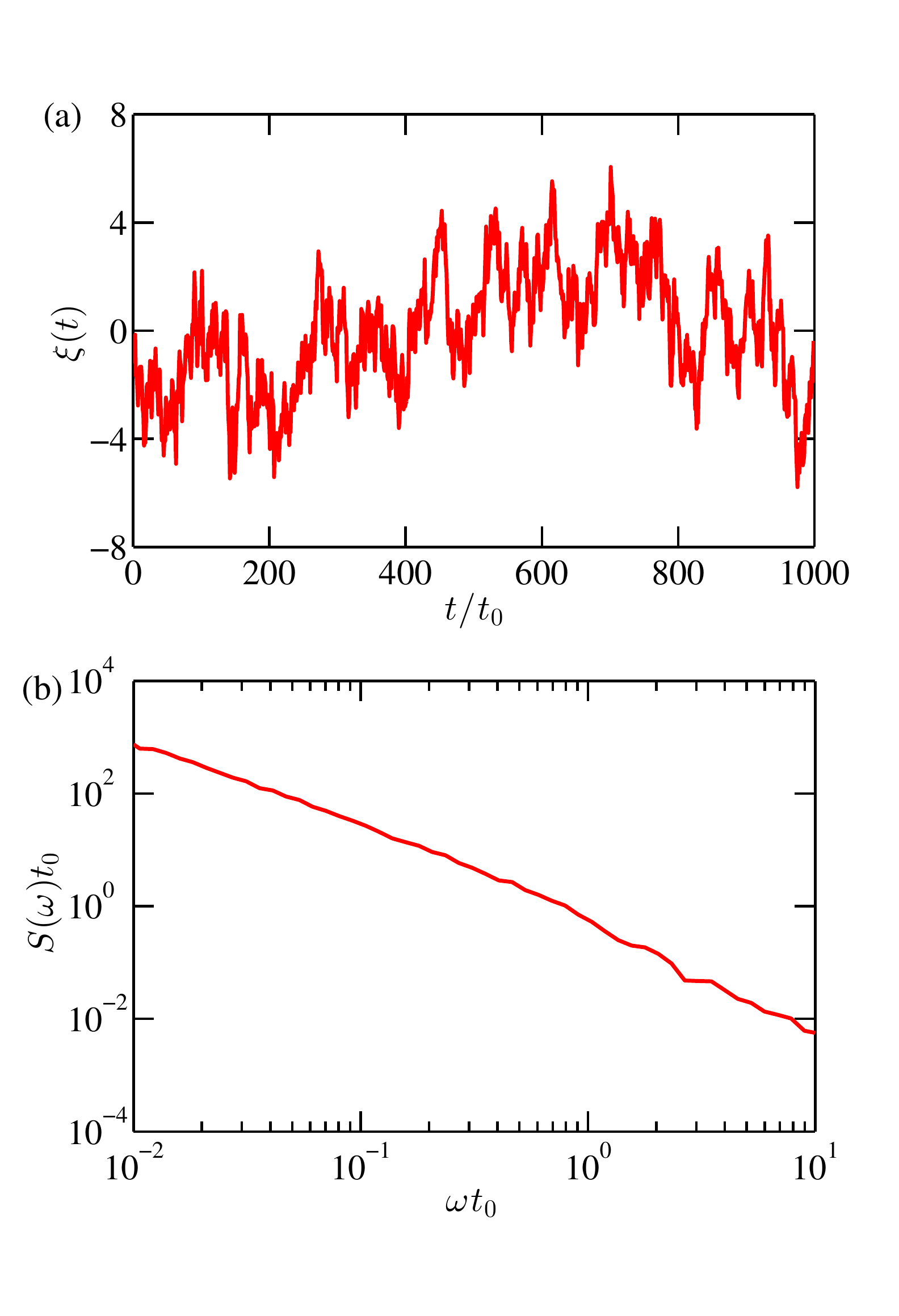}
\caption{Example of the $1/f$ noise used in the simulation. (a) Noise as a function of time.(b) The power spectral density corresponding to (a), $S(\omega)=A/(\omega t_{0})^{1.5}$. The noise amplitude $A$ has been scaled such that $S(\omega=1/t_{0})\approx 1/t_{0}$.}
\label{fig:noises example}
\end{figure}

Fig.~\ref{fig:noises example} shows an example of the time-dependent noise that we are using in this work. The $1/f$ noise has a power spectral density $S(\omega)=A/(\omega t_0)^\alpha$ where $\alpha$ is the noise exponent that determines the extent to which the noise is correlated, and $A$ is the amplitude. The $1/f$ noise is typically generated by summing random telegraph signals,\cite{Wang.14a, Wang.14b} which in principle is capable of producing noises with exponent $0\le\alpha\le2$, but in practice good convergence is only achieved for a more limited range $1/2\lesssim\alpha\lesssim3/2$. Therefore in this work we use the method described in Ref.~\onlinecite{Yang.16}, capable to generate $1/f$ noises with exponent $0\le\alpha\le3$. Fig.~\ref{fig:noises example}(a) shows $\xi(t)$, the noise as a function of time, while Fig.~\ref{fig:noises example}(b) shows the corresponding power spectral density with $\alpha=1.5$. To facilitate the discussion, we also introduce $t_0$ as the time unit throughout this work. The dimensionless control field, $J t_0$ (where $J$ is either $J_{12}$ or $J_{23}$), takes value between 0 to 1. Assuming that in experiments $J$ ranges between 100 MHz and 1 GHz, the time unit $t_0$ should be taken as $1\sim10$ ns.

\begin{figure}
	\includegraphics[width=1\columnwidth]{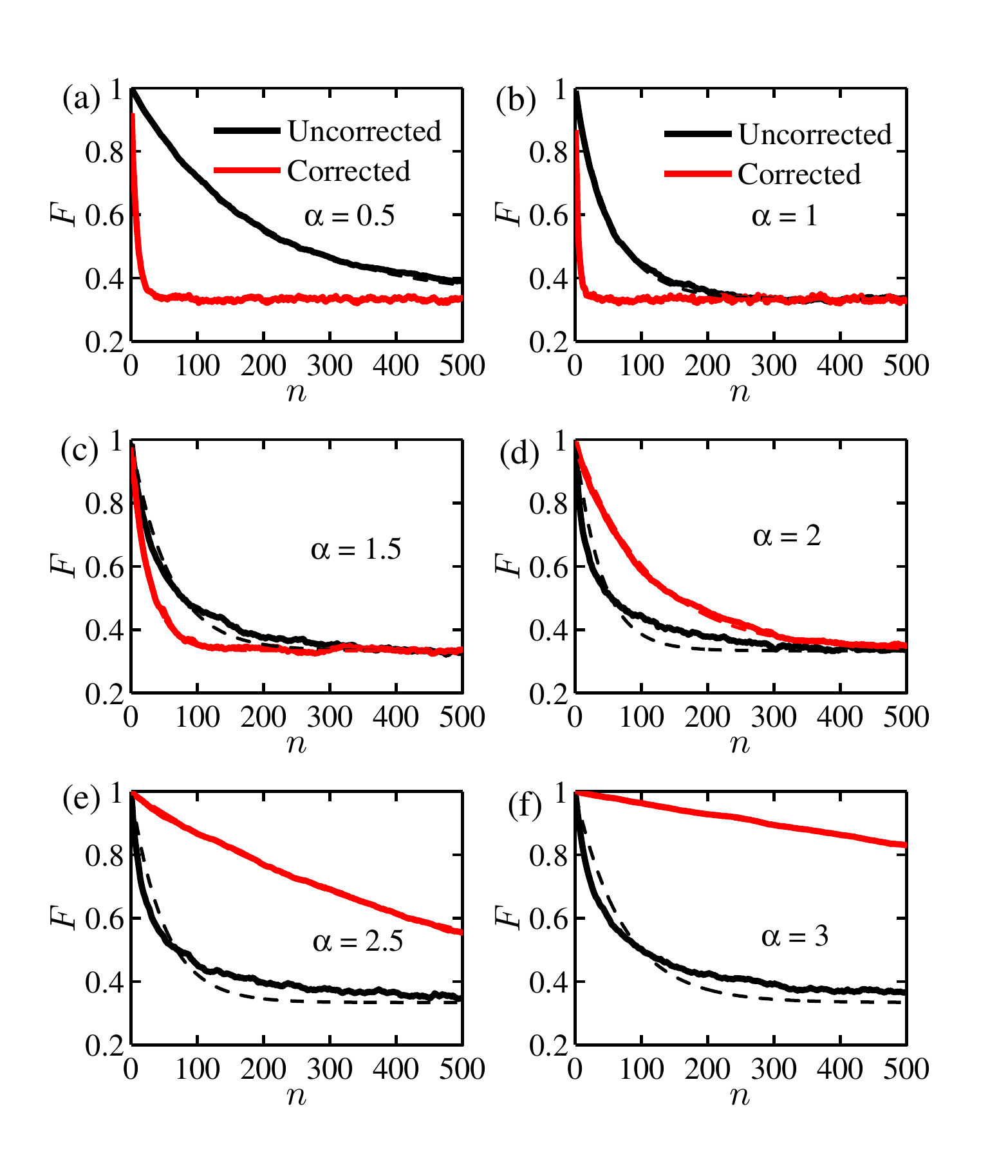}
	\caption{Randomized benchmarking of single-qubit Clifford gates undergoing $1/f$ noises with different noise exponents $\alpha$ as indicated. $n$ is the number of gates and $F$ indicates the fidelity. The results for uncorrected and corrected gates are shown as lines, respectively. The noise amplitudes for each panel are (a) $At_{0}=10^{-3}$, (b) $10^{-3}$, (c) $10^{-3.5}$, (d) $10^{-4}$, (e) $10^{-5}$, and (f) $10^{-6}$.}
	\label{fig:fidplot}
\end{figure}

While the fidelities or errors of individual quantum gates are of great interest, they cannot be straightforwardly used to predict the fidelity of a quantum algorithm, which involves a number of such gates. Moreover, when the noises are time-dependent, the duration of one quantum gate is typically too short for the noise to exhibit the desired spectrum. A widely-used approach to this problem is the Randomized Benchmarking.\cite{Knill.08,Magesan.12} The benchmarking is implemented by averaging the
gate fidelities over random sequences of gates drawn from a subset of quantum gates---the Clifford group---undergoing different realizations of noises. The average gate fidelity is then obtained by performing an exponential fit to the fidelity decay curve. This practice avoids the error introduced during the initialization and read-out, and focuses on the gate operations. In this work, we numerically simulate the Randomized Benchmarking of single-qubit Clifford gates for the exchange-only qubit under the hyperfine noise of $1/f$ type. In each run the fluctuations in $\Delta_A$ and $\Delta_B$ [cf. Eq.~\eqref{eq:Hhfdef}] are assumed to have the same spectrum but are independent. In order to ensure convergence, we have averaged the fidelities of at least 500 random gate sequences undergoing different noise realizations for a given noise spectrum before the exponential fit is made.

Fig.~\ref{fig:fidplot} shows the results of the Randomized Benchmarking of the single-qubit Clifford gates, comparing the uncorrected ones and ones corrected against the hyperfine noise. The fidelities decay approximately exponentially, and they saturate at a value of $1/3$. This is consistent with the theoretical investigation of situations involving leakage.\cite{Epstein.14} The gate error inferred from the decay of the fidelity is therefore a combination of the dephasing and leakage error. Fig.~\ref{fig:fidplot}(a) shows the case with $\alpha=0.5$, meaning that the noise is very much similar to a white noise and one does not expect that the composite pulse should offer any improvement. In fact, the steep initial drop of the red/gray line indicates that the corrected sequences have much larger error compared to the uncorrected pulses, represented by the smoothly decaying black line. In  Fig.~\ref{fig:fidplot}(b) where $\alpha=1$, it is still the case that the error is larger for the corrected sequences, but the differences from the uncorrected pulse are smaller. 

As $\alpha$ further increases, the noise concentrates more at low frequencies and the composite pulses work better, as can be seen in  Fig.~\ref{fig:fidplot}(c)-(f). To make the comparison clear, we have chosen the amplitudes for noises with different exponents such that the error for the uncorrected pulses are comparable for these panels. We note that it is difficult to make a direct comparison between noises with different $\alpha$ values because they have different energies. Therefore we show typical results for noises with specific amplitudes in Fig.~\ref{fig:fidplot}, but consider the dependences of gate error on the noise amplitudes in the next step. For $\alpha=1.5$ [Fig.~\ref{fig:fidplot}(c)], the error for uncorrected and corrected sequences are comparable, although that for the corrected sequences is still a little larger than the uncorrected ones for the amplitude $A=10^{-3.5}/t_0$. At $\alpha=2$ [Fig.~\ref{fig:fidplot}(d)] the corrected sequences start to offer improvements, and the improvements becomes much more pronounced as $\alpha$ becomes even larger. For $\alpha=2.5$ [Fig.~\ref{fig:fidplot}(e)], the gate error has been substantially suppressed by the corrected sequence so that after 500 gates the fidelity is still larger than 0.5. In the case of $\alpha=3$ [Fig.~\ref{fig:fidplot}(f)], the fidelity is maintained above 80\% after 500 gate operations, indicating the strong power of the noise-corrected sequences.

We fit the fidelity decaying curves as shown in Fig.~\ref{fig:fidplot} to
\begin{equation}
F=\frac{2}{3}e^{-\gamma n}+\frac{1}{3},\label{eq:expfit}
\end{equation}
where the fidelity decay constant $\gamma$ is related to both the dephasing and leakage error of the quantum gates. It is difficult and unnecessary to separate the two for this work, since our main focus is to investigate the improvement of the overall gate fidelity by composite pulses under different noise conditions. We will therefore use the fidelity decay constant $\gamma$ as a measure of the gate error, without distinguishing different kinds of error.

\begin{figure}
	\includegraphics[width=\columnwidth]{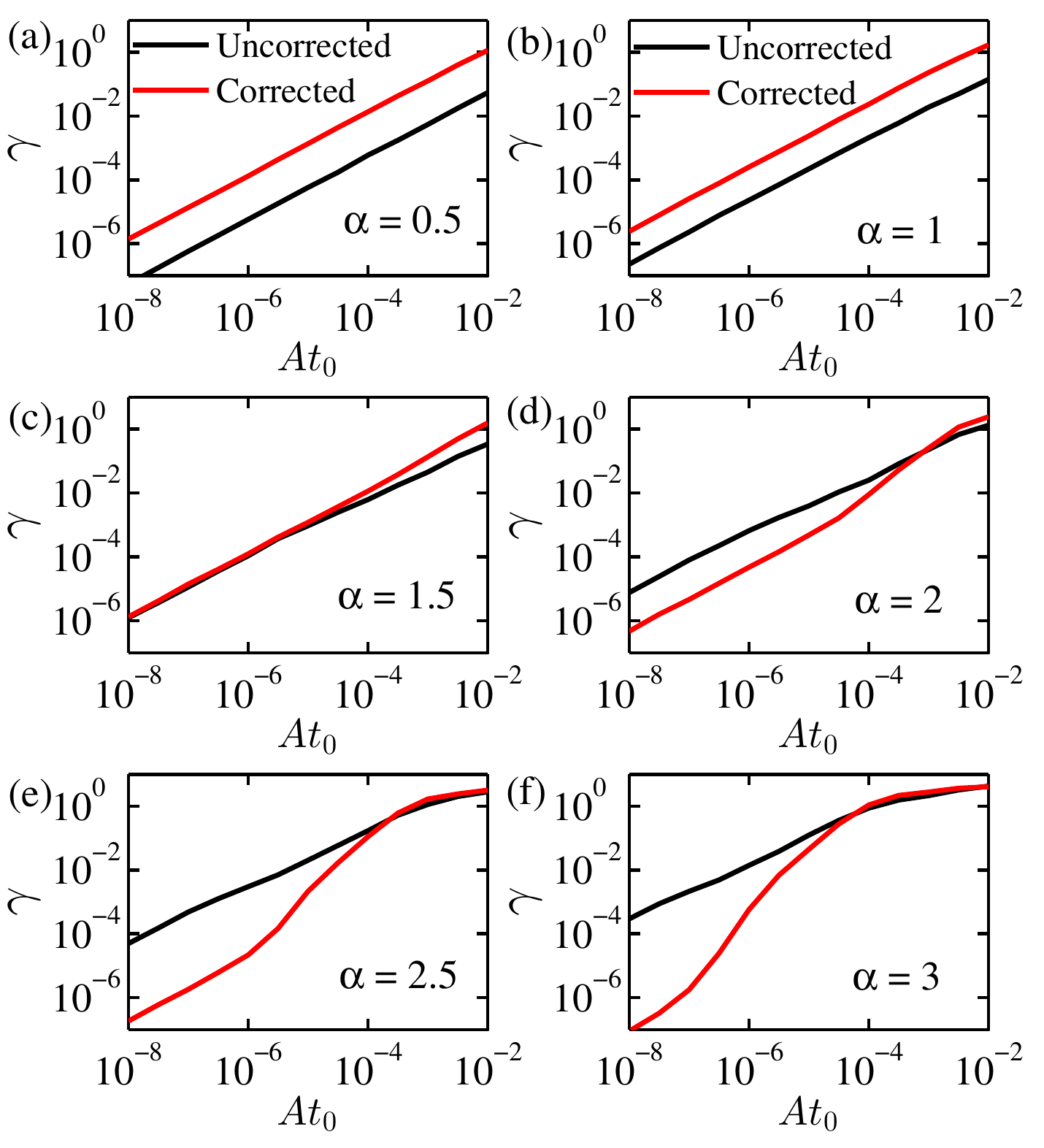}
	\caption{The fidelity decay constant $\gamma$ v.s.~ for $1/f$ noises with different noise exponents $\alpha$ as indicated. The results for uncorrected and corrected gates are shown as black and red lines, respectively. }
	\label{fig:aveerror}
\end{figure}

In Fig.~\ref{fig:aveerror} we show the dependence of the fitted fidelity decay constant $\gamma$ on the noise amplitude $A$. This can be considered as a summary of all benchmarking results, among which only representative results have been shown in Fig.~\ref{fig:fidplot}. For small $\alpha$, the noise is similar to the white noise and dynamically corrected gates will not reduce the error but rather increase it, as can be seen from Fig.~\ref{fig:aveerror}(a) and (b). For an intermediate value of $\alpha=1.5$, uncorrected and corrected pulses  have similar performances. As $\alpha$ increases, the noises become more correlated and the corrected sequences outperform the uncorrected ones. Fig.~\ref{fig:aveerror}(d) shows that for $\alpha=2$ and small enough noise $(A<10^{-4}/t_0)$ the dynamically corrected gates have reduced the error by about an order of magnitude, while in Fig.~\ref{fig:aveerror}(e), we see that the error reduction becomes about two orders of magnitude for the case of $\alpha=2.5$. The reduction is almost three orders of magnitude when $\alpha$ is as large as 3, in which case the noise becomes slowly varying and is similar to the quasi-static noise originally conceived in the development of the dynamically corrected gates. These are consistent with the qualitative consideration that DCGs work better for low frequency noises but would produce even larger errors than uncorrected pulses if the noise contains a substantial higher-frequency part.

In order to further understand how the improvement made by dynamically corrected gates relates to the structure of the noise, we define an improvement ratio $\kappa$  as the ratio between errors resulted from the uncorrected and corrected pulses under the same noise condition. We observe from Fig.~\ref{fig:aveerror} that for noises with small enough amplitudes the two curves becomes two parallel lines, and a ratio between them is well defined. The reason is that while the dynamically corrected gates can fully cancel zero frequency noise and can compensate the majority of the contributions from low frequencies, it cannot fully remove the effect of low frequency noises, let alone higher frequency parts. Therefore the first-order correction to the evolution operator due to the noises, proportional to $A$, always exist and such dependence will be made clear when $A$ is sufficiently small. Therefore the defined improvement ratio $\kappa$ is essentially the ratio between the coefficients of the leading-order error in the gate fidelity for uncorrected and corrected sequences. Note that for $\alpha=3$ [Fig.~\ref{fig:fidplot}(f)], the two lines are not parallel for $A\gtrsim10^{-7}$  and the error shows different scaling. Here, the noise varies so slowly and the DCGs works so well that compared to the second order error in the evolution operator, the first order error is almost removed  by the dynamically corrected gates. Nevertheless, for sufficiently small $A$ (in this case one must go down to $A\lesssim10^{-8}/t_0$) the two lines will still become parallel. In practice due to the lower bound to $A$ and $\gamma$ imposed by the machine precision the range that the two lines are approximately parallel is much narrower than other cases, but we have still been able to obtain a meaningful $\kappa$ for this situation. In any case, if $\kappa$ is taken before the two lines becomes parallel then the actual improvement can only be larger than that value. Therefore one may consider that such a fit may slightly underestimate the improvement afforded by the dynamically corrected gates.

\begin{figure}
  \includegraphics[width=0.9\columnwidth]{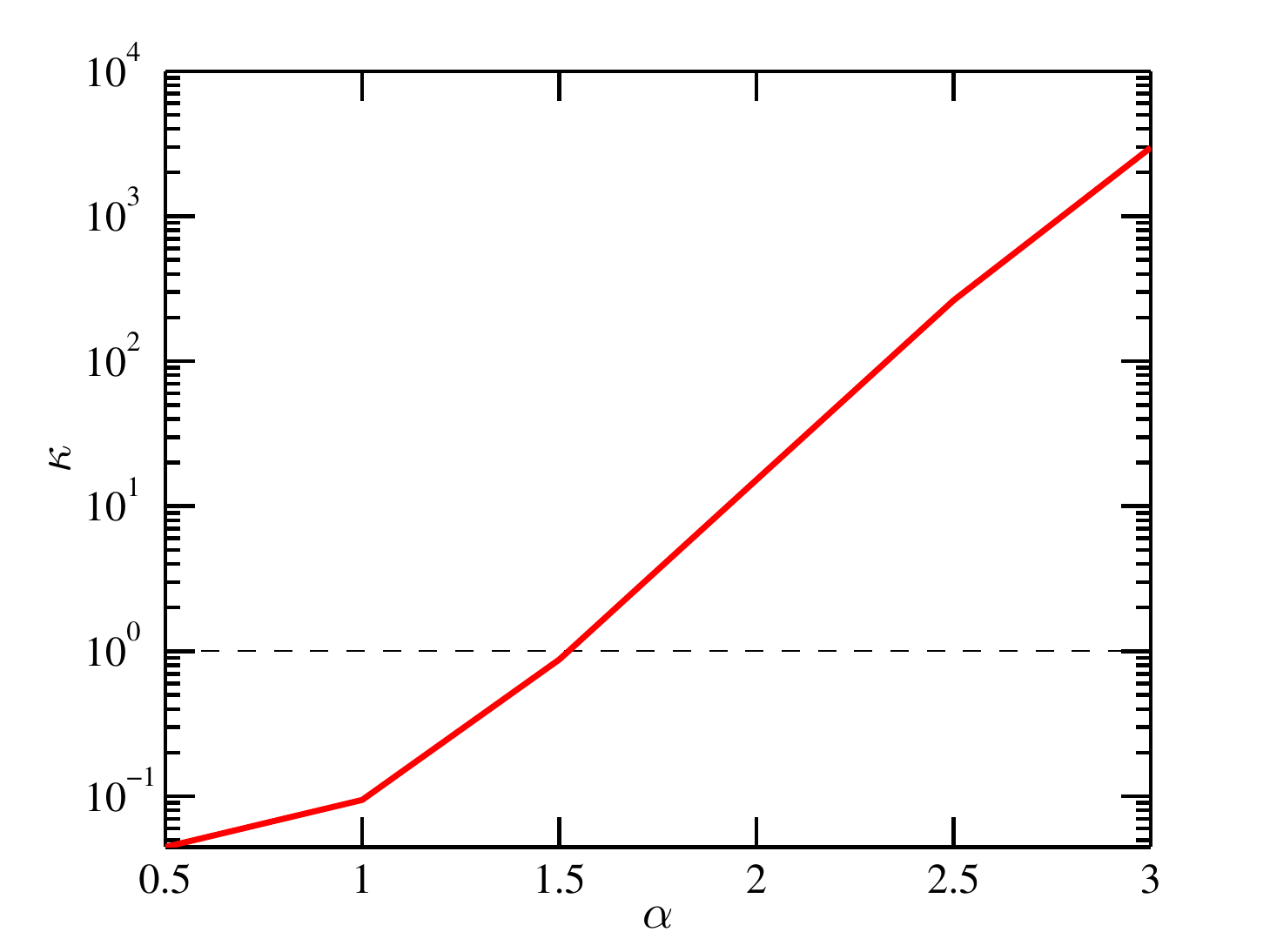}
  \caption{ The improvement ratio $\kappa$ of the dynamically corrected gates v.s. the noise exponent $\alpha$. The dashed line marks $\kappa=1$.}\label{fig:ratio}
\end{figure}

In Fig.~\ref{fig:ratio} we show the improvement ratio $\kappa$ versus the noise exponent $\alpha$. This is the core result of this paper as it provides useful information to the experimentalists who are to implement the DCGs developed in Ref.~\onlinecite{Hickman.13}. From the figure we can clearly see that for small $\alpha$ (white-like noise), $\kappa<1$, meaning that dynamically corrected gates produce larger error than the uncorrected ones due to their prolonged gate time. For large $\alpha$ (quasi-static noise) dynamically corrected gates offer great improvements as seen by the large $\kappa$ value in Fig.~\ref{fig:ratio}. It is therefore important to know the critical exponent $\alpha_c$ where the two situation cross. We have found that for the DCGs developed to combat the hyperfine noise in the exchange-only qubit,\cite{Hickman.13} $\alpha_c\approx1.5$. This value is larger than the $\alpha_c$ previously found for the singlet-triplet qubits.\cite{Yang.16} The reasons that DCGs for the exchange-only qubit requires a more correlated noise than the singlet-triplet qubit case to take effect possibly include the complexity of the pulse sequences (two control pulses for the exchange-only versus one for the singlet-triplet qubits), and most likely the presence of an additional channel of decoherence, the leakage. Fortunately, experimentally measured noise exponent for GaAs systems are still much larger than $\alpha_c\approx1.5$ (in Refs.~\onlinecite{Rudner.11,Medford.12} $\alpha$ has been measured to be $\sim2.6$).  In Ref.~\onlinecite{Medford.12}, the noise spectrum has been measured to be $S(\omega)\approx(3.6\ \mathrm{\mu s})^{-\beta-1}/\omega^\beta$ where $\beta\approx2.6$. Converting to our units it implies that $A t_0\approx(t_0/3.6\ \mathrm{\mu s})^{3.6}$. Taking $t_0=10\ \mathrm{ns}$ then $A t_0\approx10^{-9}$. Comparing to Fig.~\ref{fig:aveerror}(e) we see that this is in a range where the dynamically corrected gates offer about two orders of magnitude reduction in the error induced by the hyperfine noise. These qualitative considerations indicate that the dynamically corrected gates, proposed in Ref.~\onlinecite{Hickman.13}, are very useful to cancel hyperfine noises in actual experiments. 

\begin{figure}
	\includegraphics[width=\columnwidth]{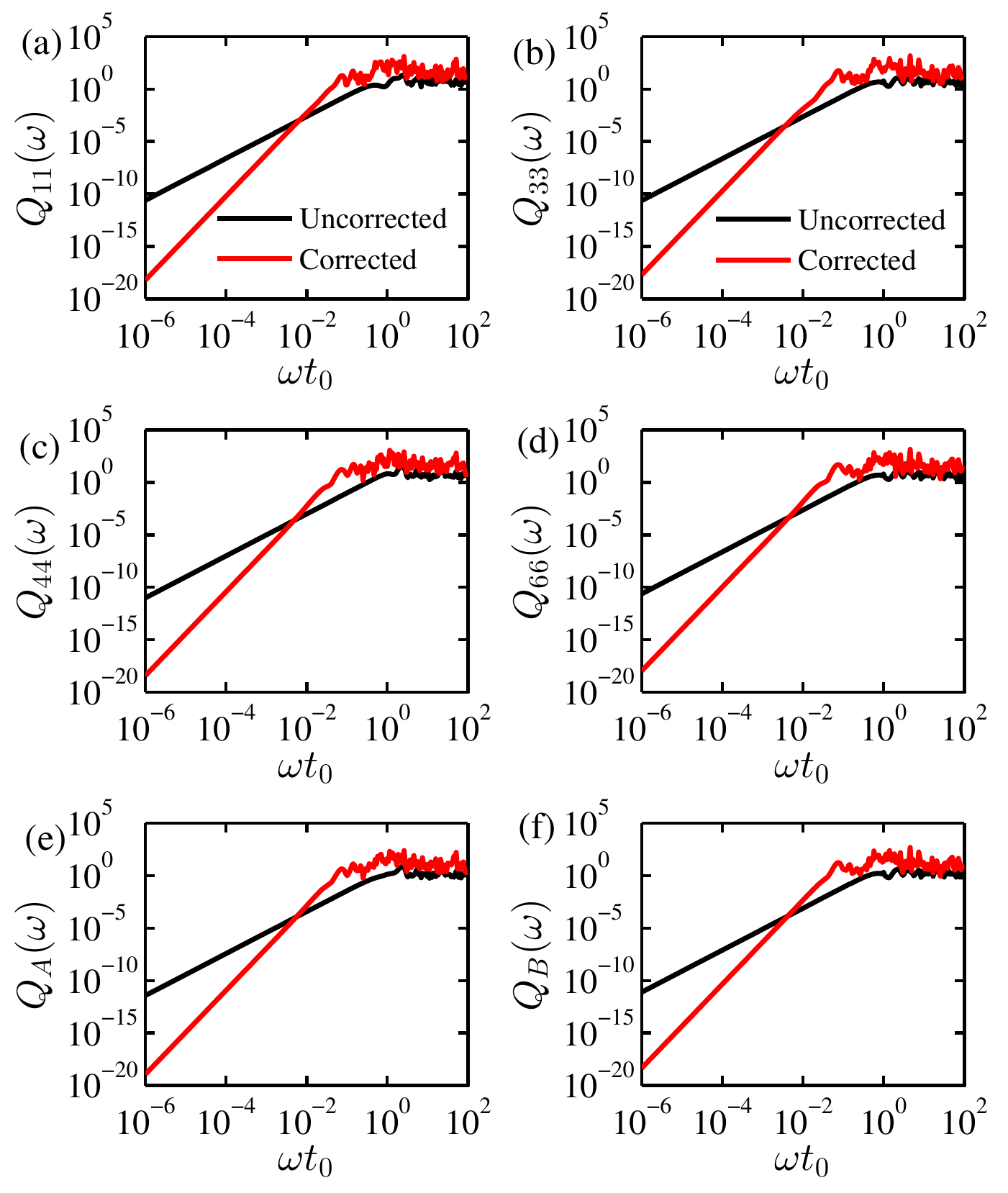}
		\caption{Filter transfer functions of the Hadamard gate $R(\hat{x}+\hat{z},\pi)$. (a)-(d): components of the filter transfer functions $Q_{11}$, $Q_{33}$, $Q_{44}$, and $Q_{66}$.(e): total filter transfer function for the noise $\Delta_A$ calculated from Eq.~\eqref{eq:filtQA}. (f): total filter transfer function for the noise $\Delta_B$ calculated from Eq.~\eqref{eq:filtQB}.}
	\label{fig:filtfunc}
\end{figure}

Another important characterization of the noise-canceling ability of quantum gates is the filter transfer function.\cite{Green.12, Green.13,Paz-Silva.14,Ball.16} 
The filter transfer function was introduced in systems involving two states and because of the presence of a leaked state in our case, the formalism needs to be slightly modified. Ref.~\onlinecite{Green.13} provides a pedagogical review of the filter function formalism and we  start from there. Eq.~(10), the control matrix, of Ref.~\onlinecite{Green.13}  should be modified in our case as
\begin{equation}
R_{ij}(t)=\mathrm{Tr}\left[U_c^\dagger(t)\lambda_iU_c(t)\lambda_j\right]/2,
\end{equation}
where $i,j$ run from 1 though 8 and $\lambda_i$'s are the Gell-Mann matrices mentioned in the previous section. The error vector, denoted by $\boldsymbol{\beta}(t)$ in Ref.~\onlinecite{Green.13}, is now an 8-dimensional vector with $\beta_1=\Delta_A/(2\sqrt{3})$,$\beta_3=\Delta_B/3$,$\beta_4=\Delta_A/\sqrt{6}$,$\beta_6=\sqrt{2}\Delta_B/3$, and all other elements being zero. The spectral density of these elements can be expressed as
\begin{equation}
S_{ij}^\beta(\omega)=\int_{-\infty}^\infty e^{-i\omega\tau}\langle\beta_i(0)\beta_j(\tau)\rangle d\tau.
\end{equation}
Using the relationship between $\beta_i$ and $\Delta_A,\Delta_B$, we may define
\begin{equation}
\begin{split}
S_{A}(\omega)&\equiv\int_{-\infty}^\infty e^{-i\omega\tau}\langle\Delta_A(0)\Delta_A(\tau)\rangle d\tau\\
&=12S_{11}^\beta(\omega)=6S_{44}^\beta(\omega),
\end{split}
\end{equation}
and
\begin{equation}
\begin{split}
S_{B}(\omega)&\equiv\int_{-\infty}^\infty e^{-i\omega\tau}\langle\Delta_B(0)\Delta_B(\tau)\rangle d\tau\\
&=9S_{33}^\beta(\omega)=\frac{9}{2}S_{66}^\beta(\omega),
\end{split}
\end{equation}
where the $A$ and $B$, as mentioned above, are used to simplify the subscripts $12$ and $\overline{12}$ in Refs.~\onlinecite{Ladd.12,Hickman.13}. With these definitions of the correlation functions of the hyperfine noises, we may write the fidelity with leading order correction as
\begin{equation}
F\simeq1-\frac{1}{2\pi}\int_{-\infty}^\infty\frac{d\omega}{\omega^2}\left[S_A(\omega)Q_A(\omega)+S_B(\omega)Q_B(\omega)\right],\label{eq:FQAQB}
\end{equation}
where $Q_A$ and $Q_B$ are the filter functions corresponding to noise channels $A$ and $B$, which can be further broken down as
\begin{equation}
Q_A(\omega)=\frac{1}{12}Q_{11}(\omega)+\frac{1}{6}Q_{44}(\omega),\label{eq:filtQA}
\end{equation}
and
\begin{equation}
Q_B(\omega)=\frac{1}{9}Q_{33}(\omega)+\frac{2}{9}Q_{66}(\omega).\label{eq:filtQB}
\end{equation}
Here, $Q_{ij}(\omega)$ is related to the Fourier transform of the control matrix [cf. Eq.~(24) in Ref.~\onlinecite{Green.13}] as
\begin{equation}
Q_{ij}(\omega)=\sum_{k=1}^8R_{ik}(\omega)R_{kj}^*(\omega).
\end{equation}
One can also readily verify that $Q_{14}$, $Q_{41}$, $Q_{36}$, and $Q_{63}$ all vanish for our problem. We note here that to a certain extent, $Q_{ij}$'s give separate characterization of dephasing and leakage errors due to the structure of the hyperfine Hamiltonian Eq.~\eqref{eq:Hhfdef}. In particular, contributions of $Q_{11}$ and $Q_{33}$ relate to dephasing, and those of $Q_{44}$ and $Q_{66}$ relate to leakage. Eq.~\eqref{eq:filtQA} and \eqref{eq:filtQB} can therefore be interpreted that contributions to the hyperfine-noise-induced gate error from leakage is about twice as large as dephasing error. Figure~\ref{fig:filtfunc} shows the results of these filter transfer functions for a chosen quantum gate, the Hadamard gate $R(\hat{x}+\hat{z},\pi)$. The filter transfer functions for other gates are very similar and are thus not shown here. The black lines show the filter transfer functions for the uncorrected sequences while red/gray ones correspond to the DCGs corrected against the hyperfine noise. The difference in scaling is apparent for all cases because the corrected pulses completely suppress leading order errors, and in realistic situations one must multiply the filter functions and the noise spectra to obtain the corrections to the fidelity as Eq.~\eqref{eq:FQAQB}. It is hard to directly compare the results between the output of Eq.~\eqref{eq:FQAQB} and the gate error derived from the Randomized Benchmarking, because the former is for individual gates while the latter is an average for a sequence of gates. Nevertheless, we have verified that the two results are of the same order of magnitude, offering a rough cross-check between the two methods. Representative results comparing the the average gate errors (the fidelity decay constant $\gamma$) for the singlet-qubit Clifford gates extracted from the Randomized Benchmarking and the errors of a specific gate, $R(\hat{x}+\hat{z},\pi)$, calculated from integrating the product of the noise spectra and the filter functions according to Eq.~\eqref{eq:FQAQB} are shown in Table~
\ref{tab:compRBfilt} for two different $\alpha$ values. We see that the errors are of the same order of magnitude for noises with the same amplitude and exponent, providing an independent check for our results of Randomized Benchmarking as well as the filter transfer functions.

\begin{table*}[t]
  \centering
\begin{tabular}{|c|c|c|c||c|c|}
\hline
  \multirow{2}{*}{$\alpha$} & \multirow{2}{*}{$At_0$} & \multicolumn{2}{|c||}{$\gamma$ from RB} & \multicolumn{2}{|c|}{error of $R(\hat{x}+\hat{z},\pi)$ from Eq.~\eqref{eq:FQAQB}} \\ \cline{3-6}
  & & uncorrected & corrected & uncorrected & corrected \\
  \hline
  \hline
   \multirow{3}{*}{$\alpha=1$} & $10^{-4}$ & $2.1\times10^{-3}$ &  $2.3\times10^{-2}$ &  $5.0\times10^{-3}$ &  $4.6\times10^{-2}$ \\
   \cline{2-6}
   & $10^{-6}$ & $2.3\times10^{-5}$ &  $2.5\times10^{-4}$ &  $5.0\times10^{-5}$ &  $4.6\times10^{-4}$ \\
    \cline{2-6}
   & $10^{-8}$ & $2.3\times10^{-7}$ &  $2.4\times10^{-6}$ &  $5.0\times10^{-7}$ &  $4.6\times10^{-6}$ \\
   \hline
   \hline
   \multirow{3}{*}{$\alpha=2$} & $10^{-4}$ & $2.5\times10^{-2}$ &  $9.0\times10^{-3}$ &  $2.5\times10^{-2}$ &  $4.4\times10^{-3}$ \\
   \cline{2-6}
   & $10^{-6}$ & $6.6\times10^{-4}$ &  $4.7\times10^{-5}$ &  $2.5\times10^{-4}$ &  $4.4\times10^{-5}$ \\
    \cline{2-6}
   & $10^{-8}$ & $7.6\times10^{-6}$ &  $4.7\times10^{-7}$ &  $2.5\times10^{-6}$ &  $4.4\times10^{-7}$ \\
   \hline
  \end{tabular}
   \caption{Comparison of the fidelity decay constant $\gamma$, related to the average gate error of single-qubit Clifford gates extracted from the Randomized Benchmarking, and the error of a specific gate, $R(\hat{x}+\hat{z},\pi)$, calculated by integrating the product of the noise spectra and the filter function according to Eq.~\eqref{eq:FQAQB}.}\label{tab:compRBfilt}
\end{table*}

\section{Conclusion}\label{sec:conclusion}

In this paper we have studied the response of dynamically corrected gates robust against the hyperfine noise in the exchange-only qubit system. The static noise approximation is essential to the theoretical development of DCGs but it is important to understand to what extent this approximation may be lifted. We simulate the noise using a $1/f$ type having a spectrum proportional to $1/\omega^\alpha$. It is as expected that for $\alpha$ small, the noise is white-like and DCGs would simply not offering any improvement but rather destruction. However, for a large value of $\alpha$ great improvement from DCGs can be observed compared to the uncorrected one. The critical $\alpha$ with which the performances of DCGs and uncorrected pulses are similar is found to be $\alpha_c\approx1.5$. This value is larger than the one for the \textsc{supcode} sequence for the singlet-triplet qubit system,\cite{Yang.16} but is still lower than the values experimentally measured for hyperfine noises,\cite{Rudner.11,Medford.12} suggesting that the DCGs developed in Ref.~\onlinecite{Hickman.13} will be very useful to suppress noise in experiments. We have also presented the filter transfer functions for the DCGs and have cross-verified that the gate error estimated from the product of the filter transfer functions and the noise spectra is similar to that obtained from the Randomized Benchmarking. Our results suggest that application of dynamically corrected gates is a useful measure to suppress noise in the system of exchange-only qubits.

This work is supported by grants from City University of Hong Kong (Projects No. 7200456 and No. 9610335).

\appendix
\renewcommand{\theequation}{A-\arabic{equation}}
\setcounter{equation}{0}
\section{single-qubit Clifford gates expressed in terms of $R_{12}$ and $R_{23}$}\label{appham}

In this Appendix we give explicit forms of the 24 single-qubit Clifford gates expressed in terms of available rotations $R_{12}(\phi)$ and $R_{23}(\phi)$.

The simplest cases are rotations around the $z$-axis, which are essentially the same as $R_{12}$ [with an additional minus sign in front of the angle due to the minus sign in Eq.~\eqref{eq:E12E23}]. The identity operator can also be expressed as a $2\pi$ rotation around $\hat{z}$. Table~\ref{tab:zrot} gives these rotations.

\begin{table}[h]
 \centering
 \begin{tabular}{|c||c|}
 \hline
 Gate & Realization  \\
 \hline
 \hline
 $R(\hat{z},-\pi/2)$ & $R_{12}(\pi/2)$\\
 \hline
  $R(\hat{z},\pi/2)$ & $R_{12}(3\pi/2)$\\
 \hline
  $R(\hat{z},\pi)$ & $R_{12}(\pi)$\\
 \hline
 $I$ & $R_{12}(2\pi)$ \\
 \hline
 \end{tabular}
 \caption{$z$-rotations and the identity operation expressed in terms of $R_{12}$ and $R_{23}$.}\label{tab:zrot}
\end{table}

For $x$-rotations, one may express $R(\hat{x},\phi)$ in terms of $R_{12}$ and $R_{23}$ according to Eq.~\eqref{eq:R23decomp}. However, attention must be paid to Eq.~\eqref{eq:psi}, because the domain of arcsin mandates that $-1\le2\sin\frac{\phi}{2}/\sqrt{3}\le1$, implying $-2\pi/3\le\phi\le2\pi/3$. For $\phi$ values outside of $[-2\pi/3,2\pi/3]$, one needs to duplicate two $x$-rotations, the angle of which lies inside the domain. For example, 
\begin{align}
R\left(\hat{x},\frac{\pi}{2}\right)=R_{12}\left[-\phi_a'\left(\frac{\pi}{2}\right)\right]R_{23}\left[\psi\left(\frac{\pi}{2}\right)\right]R_{12}\left[-\phi_a'\left(\frac{\pi}{2}\right)\right],
\end{align}
and $R(\hat{x},\pi)$ is realized by two consecutive $R(\hat{x},\pi/2)$ rotations. For the convenience of discussion, we define $\phi_a'\left(\pi/2\right)=\arctan\sqrt{1/2}\equiv\eta$. and $-\psi\left(\pi/2\right)=2\arcsin\sqrt{2/3}\equiv\xi$. These angles frequently appear in the tables below. Table~\ref{tab:xrot} gives how $x$-rotations are decomposed into combinations of $R_{12}$ and $R_{23}$.

\begin{table*}[h]
 \centering
 \begin{tabular}{|c||c|}
 \hline
 Gate & Realization  \\
 \hline
 \hline
 $R(\hat{x},-\pi/2)$ & $R_{12}\left(\eta\right)R_{23}\left(\xi\right)R_{12}\left(\eta\right)$\\
 \hline
  $R(\hat{x},\pi/2)$ & $R_{12}\left(-\eta\right)R_{23}\left(-\xi\right)R_{12}\left(-\eta\right)$\\
 \hline
  $R(\hat{x},\pi)$ & $R_{12}\left(-\eta\right)R_{23}\left(-\xi\right)R_{12}\left(-2\eta\right)R_{23}\left(-\xi\right)R_{12}\left(-\eta\right)$\\
 \hline
 \end{tabular}
 \caption{$x$-rotations expressed in terms of $R_{12}$ and $R_{23}$.}\label{tab:xrot}
\end{table*}

For other rotations, one first decompose it into an $x$-rotation sandwiched by two $z$-rotations, and then use the decomposition of $x$-rotation in terms of $R_{12}$ and $R_{23}$ as Eq.~\eqref{eq:R23res}. Adjacent $R_{12}$ operations can be combined in order to optimize the total gate time. The results are shown in Table.~\ref{tab:genrot}

\begin{table*}[h]
	\centering
	\begin{tabular}{|c||c|}
		\hline
		Gate & Realization  \\
		\hline
		\hline
		$R(\hat{y},-\pi/2)$ & $R_{12}\left(\eta+3\pi/2\right)R_{23}\left(\xi\right)R_{12}\left(\eta+\pi/2\right)$\\
		\hline
		$R(\hat{y},\pi/2)$ &$R_{12}\left(-\eta+3\pi/2\right)R_{23}\left(-\xi\right)R_{12}\left(-\eta+\pi/2\right)$\\
		\hline
		$R(\hat{y},\pi)$ & $R_{12}\left(-\eta+3\pi/2\right)R_{23}\left(-\xi\right)R_{12}\left(-2\eta\right)R_{23}\left(-\xi\right)R_{12}\left(-\eta+\pi/2\right)$\\
		\hline
		$R(\hat{x}+\hat{z},\pi)$ &$R_{12}\left(\eta+\pi/2\right)R_{23}\left(\xi\right)R_{12}\left(\eta+\pi/2\right)$\\
		\hline
		$R(\hat{x}-\hat{z},\pi)$ &$R_{12}\left(\eta+3\pi/2\right)R_{23}\left(\xi\right)R_{12}\left(\eta+3\pi/2\right)$\\
		\hline
		$R(\hat{x}+\hat{y},\pi)$ &$R_{12}\left(\eta+\pi/2\right)R_{23}\left(\xi\right)R_{12}\left(2\eta\right)R_{23}\left(\xi\right)R_{12}\left(\eta+\pi\right)$\\
		\hline
		$R(\hat{x}-\hat{y},\pi)$ &$R_{12}\left(\eta+\pi/2\right)R_{23}\left(\xi\right)R_{12}\left(2\eta\right)R_{23}\left(\xi\right)R_{12}\left(\eta\right)$\\
		\hline
		$R(\hat{y}+\hat{z},\pi)$ &$R_{12}\left(\eta\right)R_{23}\left(\xi\right)R_{12}\left(\eta+\pi\right)$\\
		\hline
		$R(\hat{y}-\hat{z},\pi)$ &$R_{12}\left(\eta+\pi\right)R_{23}\left(\xi\right)R_{12}\left(\eta\right)$\\
		\hline
		$R(\hat{x}+\hat{y}+\hat{z},2\pi/3)$ &$R_{12}\left(-\eta+3\pi/2\right)R_{23}\left(-\xi\right)R_{12}\left(-\eta\right)$\\
		\hline
		$R(\hat{x}+\hat{y}+\hat{z},4\pi/3)$ &$R_{12}\left(-\eta+\pi\right)R_{23}\left(-\xi\right)R_{12}\left(-\eta+3\pi/2\right)$\\
		\hline
		$R(\hat{x}+\hat{y}-\hat{z},2\pi/3)$ &$R_{12}\left(-\eta\right)R_{23}\left(-\xi\right)R_{12}\left(-\eta+\pi/2\right)$\\
		\hline
		$R(\hat{x}+\hat{y}-\hat{z},4\pi/3)$ &$R_{12}\left(-\eta+\pi/2\right)R_{23}\left(-\xi\right)R_{12}\left(-\eta+\pi\right)$\\
		\hline
		$R(\hat{x}-\hat{y}+\hat{z},2\pi/3)$ &$R_{12}\left(-\eta\right)R_{23}\left(-\xi\right)R_{12}\left(-\eta+3\pi/2\right)$\\
		\hline
		$R(\hat{x}-\hat{y}+\hat{z},4\pi/3)$ &$R_{12}\left(-\eta+3\pi/2\right)R_{23}\left(-\xi\right)R_{12}\left(-\eta+\pi\right)$\\
		\hline
		$R(-\hat{x}+\hat{y}+\hat{z},2\pi/3)$ &$R_{12}\left(-\eta+\pi\right)R_{23}\left(-\xi\right)R_{12}\left(-\eta+\pi/2\right)$\\
		\hline
		$R(-\hat{x}+\hat{y}+\hat{z},4\pi/3)$ &$R_{12}\left(-\eta+\pi/2\right)R_{23}\left(-\xi\right)R_{12}\left(-\eta\right)$\\
		\hline
	\end{tabular}
	\caption{Rotations other than $z$- and $x$-rotations expressed in terms of $R_{12}$ and $R_{23}$.}\label{tab:genrot}
\end{table*}

Before we end this section, we note that in practice when one replaces for example $R_{12}(\phi)$ by $U_{12}(\phi')$, care must be exercised  in regards to the domain of $\phi'$ (between $-\pi$ and $\pi$) so that $2\pi$'s may be required to be added or subtracted to or from $\phi$.

%

\end{document}